\documentclass[conference,a4paper]{IEEEtran}
\IEEEoverridecommandlockouts
\usepackage{cite}
\usepackage{tikz}
\usetikzlibrary{arrows.meta, positioning, calc}
\usepackage{amsmath, amssymb}
\usepackage[T1]{fontenc}
\usepackage{helvet}
\usepackage{amsmath,amssymb,amsfonts}
\usepackage{algorithm}
\usepackage{algorithmic}
\usepackage{graphicx}
\usepackage{textcomp}
\usepackage{xcolor}
\usepackage[utf8]{inputenc}
\usepackage[T1]{fontenc}
\usepackage{cite}
\usepackage{amsmath,amssymb,amsfonts}
\usepackage{algorithmic}
\usepackage{graphicx}
\usepackage{textcomp}
\usepackage{xcolor}
\usepackage{booktabs} 
\usepackage{multirow} 
\usepackage{url}      
\usepackage{bm}       
\usepackage{algorithm}      
\usepackage{algorithmic}    
\usepackage{amsmath}       
\usepackage{amssymb}       
\usepackage{tikz}
\usetikzlibrary{shapes.geometric, arrows.meta, positioning, fit, backgrounds}
\usepackage{algorithm}      
\usepackage{algorithmic}    

\usepackage{booktabs}       
\usepackage{graphicx}       

\usepackage{amssymb}        
\def\BibTeX{{\rm B\kern-.05em{\sc i\kern-.025em b}\kern-.08em
    T\kern-.1667em\lower.7ex\hbox{E}\kern-.125emX}}
\begin{document}

\title{Trust-Aware Multi-Agent Traceability: Confidence-Calibrated Knowledge Graphs for Consistent Software Artifact Management}

\author{\IEEEauthorblockN{Mohamed Essam, Kareem Wael, Azza Hassan, Ahmed Haitham, Mahmoud Soliman*, Samer Saber, Ibrahim Habib}
\IEEEauthorblockA{\textit{CairoMotive} \\
Cairo, Egypt \\
\textsuperscript{*}mahmoud.soliman@cairomotive.com}
}


\maketitle

\begin{abstract}
Multi-agent AI systems are increasingly used to automate software engineering tasks including requirements analysis, architecture design, test generation, and traceability linking. When these agents operate as a sequential pipeline over shared software artifacts, errors and low-confidence decisions made by upstream agents propagate to downstream stages, producing orphaned requirements, contradictory links, and compliance gaps that pose significant risks in safety-critical domains. We propose a trust-aware coordination framework where a shared knowledge graph serves as both centralized semantic memory and a coordination surface through which agents assess and build upon each other's contributions using calibrated confidence scores. Our approach introduces a two-stage traceability link prediction pipeline combining embedding-based retrieval with LLM-based multi-criteria analysis, a traceability seeding mechanism that enables comparison between derivation-time and validation-time confidence, and a consistency protocol governing pipeline interactions through confidence threshold gating, confidence divergence detection, and conflict resolution. We evaluate on an automotive software engineering case study measuring link prediction calibration, protocol effectiveness, threshold sensitivity, and the impact of traceability seeding. Ablation studies confirm that confidence calibration is essential for effective pipeline coordination.
\end{abstract}


\section{Introduction}

Modern software development increasingly relies on AI agents to automate tasks across the software lifecycle, from requirements elicitation and architectural design to code generation and test synthesis~\cite{fan2023large, jimenez2024swebench}. As these agents grow more capable, a natural evolution is deploying multiple specialized agents that collaborate on different facets of the same project. A requirements agent might decompose stakeholder needs into structured specifications, while a component agent designs the software architecture and a test agent generates verification procedures. This multi-agent paradigm promises significant productivity gains and has attracted growing research interest~\cite{hong2024metagpt, qian2024chatdev}.

However, a fundamental coordination challenge emerges when these agents operate as a sequential pipeline over shared software artifacts. A requirements agent derives software requirements and seeds initial traceability links, a component agent designs the architecture, a traceability agent validates and extends linkage across the full artifact space, and a test agent generates verification procedures building on the established links. In this pipeline, errors and low-confidence decisions made by upstream agents propagate to downstream stages: poorly seeded traceability links lead to misaligned components, which in turn produce inadequate tests. Without a mechanism for downstream agents to assess the reliability of upstream contributions, inconsistencies accumulate, including orphaned requirements, contradictory links, and compliance gaps. These problems are especially consequential in safety-critical domains, where standards such as ISO 26262~\cite{iso26262} and ASPICE~\cite{aspice2023} mandate complete and verifiable traceability across all software artifacts.

Existing approaches address parts of this challenge in isolation. Software traceability recovery has been studied extensively using information retrieval techniques~\cite{borg2014recovering, guo2017semantically}, and more recently through deep learning and transformer-based models~\cite{lin2021traceability, rodriguez2023prompts}. Knowledge graphs have been proposed as structured representations for software artifacts, enabling richer queries and relationship modeling~\cite{zhao2023software, rath2018traceability}. Multi-agent systems have been applied to collaborative software engineering tasks~\cite{hong2024metagpt, qian2024chatdev}, though primarily for code generation rather than cross-lifecycle artifact management. However, no existing work addresses the intersection of these concerns: how multiple AI agents should coordinate their contributions to a shared traceability knowledge graph, and how the reliability of individual agent decisions should influence downstream agent behavior.

We argue that the missing element is \textit{trust-aware coordination}. When an upstream agent creates a traceability link, the confidence it assigns should not be merely a diagnostic byproduct; it should serve as a first-class signal that downstream agents consume when deciding whether to build upon, flag for review, or reject that link. Furthermore, when a validation agent independently reassesses a link seeded by a derivation agent, the divergence between their confidence scores becomes a structured signal for detecting inconsistencies. This 
reframes the shared knowledge graph from passive storage into an active coordination surface where pipeline stages communicate trust through calibrated confidence.

In this paper, we propose a framework that addresses both the quality of individual traceability predictions and the consistency of multi-agent interactions. Our contributions are as follows:

\begin{enumerate}
    \item A two-stage traceability link prediction pipeline that combines embedding-based semantic retrieval with LLM-based multi-criteria analysis to produce confidence-calibrated links between software artifacts, evaluated using calibration-aware metrics including Expected Calibration Error and Brier score.
    
    \item A traceability seeding mechanism in which derivation agents produce initial links with self-assessed confidence, and a validation agent independently reassesses these links, enabling structured comparison between derivation-time and validation-time confidence scores.
    
    \item A consistency protocol for pipeline-based multi-agent knowledge graph operations comprising three mechanisms: confidence threshold gating, which prevents downstream agents from building on low-confidence links; confidence divergence detection, which identifies disagreements between seeding and validation assessments; and conflict resolution, which materializes detected inconsistencies as first-class graph entities for human review.
    
    \item An empirical evaluation on an automotive software engineering case study, including ablation studies on link prediction accuracy, calibration quality, protocol effectiveness, threshold sensitivity, and the impact of traceability seeding.
\end{enumerate}
 \section{Related Work}
  \label{sec:related}

  Our work sits at the intersection of four research areas: software traceability recovery, knowledge graphs for software
  engineering, multi-agent pipelines for collaborative development, and confidence calibration in language models.

  \subsection{Software Traceability Recovery}

  Automated traceability link recovery has evolved from information retrieval techniques (VSM,
  LSI)~\cite{antoniol2002recovering, borg2014recovering} to deep learning methods. Word
  embeddings~\cite{guo2017semantically} and transformer-based architectures~\cite{lin2021traceability} substantially
  outperform lexical matching. Recent approaches include LLM-based prompt engineering~\cite{rodriguez2023prompts} and
  retrieval-augmented generation (RAG) pipelines~\cite{hey2024rag}. In safety-critical domains, LLM-based embedding
  similarity for MBSE artifacts~\cite{bonner2024llm} and LLM-assisted entity extraction~\cite{keim2025architecture} have
  shown promise.

  However, existing approaches share two critical gaps: First, they produce link predictions without calibrated confidence
   estimates that downstream processes can meaningfully consume. Second, they do not distinguish between traceability
  links established during artifact derivation versus those recovered through post-hoc validation. Our work exploits this
  dual-confidence signal by enabling structured comparison between derivation-time and validation-time assessments.

  \subsection{Knowledge Graphs for Software Engineering}

  Knowledge graphs effectively represent software artifact relationships~\cite{zhao2023software, rath2018traceability},
  enabling efficient impact analysis and gap detection~\cite{clelandhuang2014software}. Recent multi-agent systems like
  AGENTiGraph~\cite{agenticgraph2024} and KARMA~\cite{karma2025} integrate LLM-based agents with graph management.
  However, these focus on general-purpose knowledge management rather than the specific consistency requirements of
  regulated software traceability.

  Our framework differs by treating the knowledge graph as an active coordination surface where calibrated confidence
  scores mediate inter-agent trust and decision-making.

  \subsection{Multi-Agent Systems for Software Engineering}

  MetaGPT~\cite{hong2024metagpt} and ChatDev~\cite{qian2024chatdev} demonstrate viability of multi-agent software
  development with role specialization. However, they focus on code generation workflows and lack mechanisms for
  downstream agents to assess reliability of upstream outputs. Trust-based coordination exists in distributed
  AI~\cite{jennings2000agent, wooldridge1995intelligent}, but assumes concurrent peer interactions rather than sequential
  pipeline propagation.

  \subsection{Confidence Calibration in Language Models}

  Verbalized confidence scores from LLMs can be well-calibrated~\cite{tian2023calibration}. Post-hoc calibration
  techniques (temperature scaling, Platt scaling)~\cite{guo2017calibration} address overconfidence, and prompting
  strategies significantly affect calibration quality~\cite{xiong2024llmuncertainty}.

  However, prior work examines individual model outputs for isolated tasks. No work has examined how calibrated confidence
   scores from different pipeline stages can be compared to detect inconsistencies. Our framework demonstrates that
  confidence calibration is essential for detecting and resolving cross-stage inconsistencies.

  \subsection{Summary and Positioning}

  Table~\ref{tab:related_comparison} summarizes positioning relative to existing approaches.

  \begin{table}[t]
  \centering
  \footnotesize
  \caption{Comparison with related work across key dimensions.}
  \label{tab:related_comparison}
  \begin{tabular}{lccccc}
  \hline
  \textbf{Approach} & \textbf{Trace.} & \textbf{KG} &
  \textbf{Pipeline} & \textbf{Calib.} & \textbf{Seed/Val.} \\
  \hline
  IR-based TLR~\cite{borg2014recovering} & \checkmark & & & & \\
  BERT-based TLR~\cite{lin2021traceability} & \checkmark & & & & \\
  LLM-based TLR~\cite{rodriguez2023prompts} & \checkmark & & & & \\
  Software KGs~\cite{zhao2023software} & \checkmark & \checkmark & & & \\
  AGENTiGraph~\cite{agenticgraph2024} & & \checkmark & \checkmark & & \\
  KARMA~\cite{karma2025} & & \checkmark & \checkmark & & \\
  MetaGPT~\cite{hong2024metagpt} & & & \checkmark & & \\
  ChatDev~\cite{qian2024chatdev} & & & \checkmark & & \\
  LLM Calibration~\cite{tian2023calibration} & & & & \checkmark & \\
  \textbf{Ours} & \checkmark & \checkmark & \checkmark & \checkmark & \checkmark \\
  \hline
  \end{tabular}
  \end{table}

  No existing work simultaneously addresses software traceability, knowledge graph-based artifact management,
  pipeline-based multi-agent coordination, confidence calibration, and dual-phase seeding and validation. Our framework
  uniquely combines all five dimensions, using calibrated confidence as the mechanism through which pipeline consistency
  is achieved.
\section{Approach}
\label{sec:approach}
\begin{figure*}[t]
\centering
\begin{tikzpicture}[
    node distance=0.6cm and 0.8cm,
    box/.style={rectangle, draw, rounded corners=3pt, minimum height=0.8cm, minimum width=2.0cm, align=center, font=\footnotesize},
    agent/.style={box, fill=blue!15, minimum width=1.7cm},
    service/.style={box, fill=orange!15, minimum width=1.9cm},
    protocol/.style={box, fill=green!12, minimum width=2.2cm},
    db/.style={cylinder, draw, shape border rotate=90, aspect=0.25, minimum height=1.0cm, minimum width=1.8cm, align=center, font=\footnotesize, fill=purple!10},
    groupbox/.style={rectangle, draw, dashed, rounded corners=5pt, inner sep=8pt},
    arr/.style={-{Stealth[length=2mm]}, thick},
    darr/.style={-{Stealth[length=2mm]}, thick, dashed},
    pipelinearr/.style={-{Stealth[length=2mm]}, thick, color=blue!60},
    label/.style={font=\footnotesize\bfseries},
]

\node[agent] (req_agent) {Requirement\\Agent};
\node[agent, right=0.4cm of req_agent] (arch_agent) {Architect\\Agent};
\node[agent, right=0.4cm of arch_agent] (trace_agent) {Traceability\\Agent};
\node[agent, right=0.4cm of trace_agent] (test_agent) {Test\\Agent};
\node[agent, right=0.4cm of test_agent] (analysis_agent) {Analysis\\Agent};

\draw[pipelinearr] (req_agent.east) -- (arch_agent.west);
\draw[pipelinearr] (arch_agent.east) -- (trace_agent.west);
\draw[pipelinearr] (trace_agent.east) -- (test_agent.west);
\draw[pipelinearr] (test_agent.east) -- (analysis_agent.west);

\node[groupbox, fit=(req_agent)(arch_agent)(trace_agent)(test_agent)(analysis_agent), label={[label, anchor=south]above:Agent Pipeline $\Pi$}] (agent_group) {};

\node[protocol, below=1.0cm of trace_agent, minimum width=9.0cm] (protocol_box) {Consistency Protocol: Confidence Gating $\mid$ Divergence Detection $\mid$ Conflict Materialization};

\node[service, below=1.0cm of protocol_box, xshift=-3.5cm] (req_svc) {Requirement\\Service};
\node[service, right=0.4cm of req_svc] (comp_svc) {Component\\Service};
\node[service, right=0.4cm of comp_svc] (trace_svc) {Traceability\\Service};
\node[service, right=0.4cm of trace_svc] (embed_svc) {Embedding\\Service};

\node[groupbox, fit=(req_svc)(comp_svc)(trace_svc)(embed_svc), label={[label, anchor=south]above:Graph Service (Orchestrator)}] (svc_group) {};

\node[box, fill=yellow!15, below=1.0cm of svc_group, minimum width=9.0cm] (pipeline) {Link Prediction: Semantic Retrieval $\rightarrow$ LLM Multi-Criteria Analysis $\rightarrow$ Calibration};

\node[db, below=1.0cm of pipeline] (neo4j) {Neo4j\\Knowledge\\Graph};

\draw[arr] (agent_group.south) -- (protocol_box.north);

\draw[arr] (protocol_box.south) -- (svc_group.north);

\draw[arr] (svc_group.south) -- (pipeline.north);

\draw[arr] (pipeline.south) -- (neo4j.north);

\coordinate (rw_v_x) at ([xshift=1.2cm]embed_svc.east);
\coordinate (darr_v_x) at ([xshift=2.4cm]embed_svc.east);

\draw[{Stealth[length=2mm]}-{Stealth[length=2mm]}, thick]
  (neo4j.east) -- (neo4j.east -| rw_v_x) coordinate (rw_c)
  -- (rw_c |- embed_svc.east)
  -- (embed_svc.east) node[pos=0.25, above, font=\scriptsize] {read/write};

\draw[darr] (pipeline.east) -- (pipeline.east -| darr_v_x) coordinate (tmp) 
  -- (tmp |- protocol_box.east) node[midway, left, font=\scriptsize] {$\hat{c},\; c_{\text{seed}},\; c_{\text{val}}$} 
  -- (protocol_box.east);

\draw[darr, color=red!50] (req_agent.south west) -- ++(0,-0.35) node[below, font=\scriptsize, color=red!70, align=center] {seeds $c_{\text{seed}}$};

\end{tikzpicture}
\caption{Architecture of the trust-aware multi-agent traceability framework. Agents operate as an ordered pipeline $\Pi$ where each stage reads from and writes to the shared knowledge graph through a Graph Service orchestrator. The Requirement Agent seeds initial traceability links with derivation confidence $c_{\text{seed}}$, which the Traceability Agent later validates producing $c_{\text{val}}$. The consistency protocol applies confidence threshold gating, divergence detection between $c_{\text{seed}}$ and $c_{\text{val}}$, and conflict materialization. Calibrated confidence scores flow back to the protocol to govern downstream agent behavior.}
\label{fig:architecture}
\end{figure*}
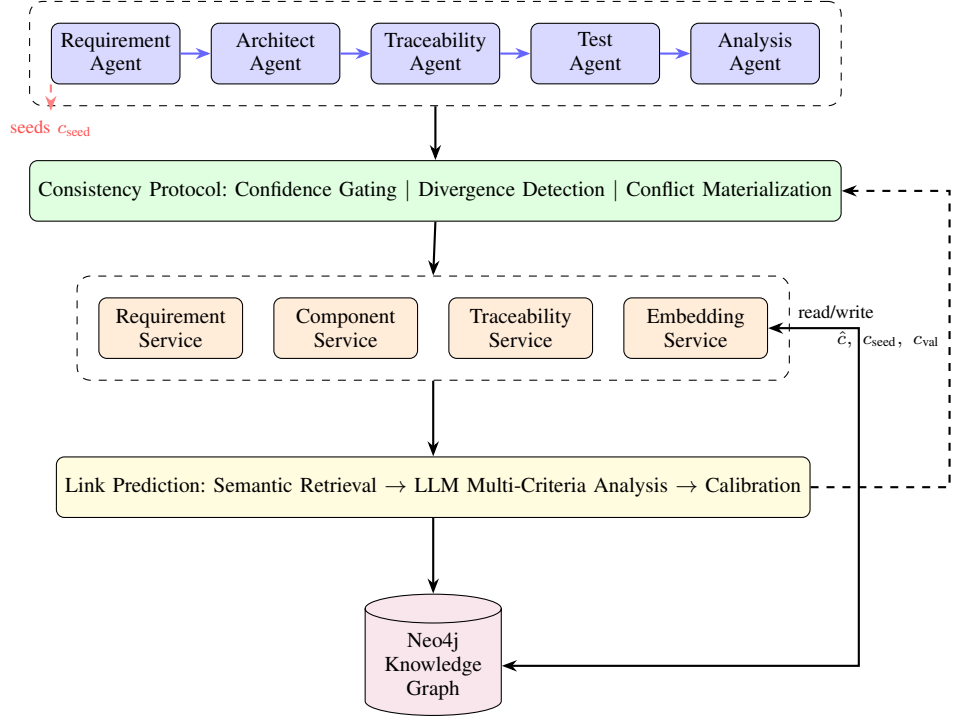

This section presents the proposed architecture for knowledge-graph-based artifact management and automated traceability generation. Fig.~\ref{fig:architecture} provides an overview of the system.

\subsection{Knowledge Graph Schema}

The knowledge graph represents engineering artifacts as nodes and traceability links as directed edges. We model this as a labeled property graph $\mathcal{G} = (\mathcal{V}, \mathcal{E}, \phi, \psi)$, where $\mathcal{V}$ is the set of nodes, $\mathcal{E} \subseteq \mathcal{V} \times \mathcal{V}$ is the set of directed edges, $\phi: \mathcal{V} \rightarrow \mathcal{L}_V$ assigns node types, and $\psi: \mathcal{E} \rightarrow \mathcal{L}_E$ assigns edge types. The node and edge type sets are:
\begin{equation}
\mathcal{L}_V = \{\texttt{Req}, \texttt{Comp}, \texttt{Test}, \texttt{Conflict}\},
\end{equation}
\begin{equation}
\mathcal{L}_E = \{\texttt{IMP\_BY}, \texttt{REF\_BY}, \texttt{VER\_BY}, \texttt{AFFECTS}\}.
\end{equation}

\noindent where \texttt{Req} encompasses both system requirements (SYS.2) and software requirements (SWE.1), \texttt{Comp} represents software components (SWE.2), \texttt{Test} represents test specifications (SWE.6), and \texttt{Conflict} represents detected inconsistencies. The edge types encode traceability relationships: \texttt{IMP\_BY} links requirements to implementing components, \texttt{REF\_BY} links software requirements to their originating system requirements, \texttt{VER\_BY} links requirements to test specifications, and \texttt{AFFECTS} connects conflict nodes to the artifacts they concern.

Each node $v \in \mathcal{V}$ carries a textual description $\text{desc}(v)$ and a dense embedding vector $\mathbf{e}_v \in \mathbb{R}^d$ representing its semantic content. Each edge $e \in \mathcal{E}$ carries a confidence score $c(e) \in [0, 1]$, a natural language justification $j(e)$, the identifier of the creating agent $\alpha(e)$, and a status $s(e) \in \{\texttt{PENDING}, \texttt{VALIDATED}, \texttt{CONFLICT}\}$.

\subsection{Layered Graph Service Architecture}

The system follows a layered architecture that separates database access from business logic. At the lowest level, the Database Layer manages the Neo4j connection, schema constraints, and vector indexes for embedding-based retrieval. Above it, the Repository Layer provides direct database access through Cypher queries for artifact creation, retrieval, and traceability operations. The Service Layer implements business logic including artifact creation, embedding generation, traceability validation, and confidence scoring. These services are coordinated by the Graph Service, which acts as a fa\c{c}ade orchestrating interactions between repositories and services. At the top, specialized AI agents interact with the knowledge graph exclusively through the Graph Service, ensuring that all operations pass through a single controlled interface.

\subsection{Agent Pipeline}

The agents operate as an ordered pipeline where each stage reads from and writes to the shared graph $\mathcal{G}$:
\begin{equation}
\Pi = A_{\text{req}} \rightarrow A_{\text{arch}} \rightarrow A_{\text{trace}} \rightarrow A_{\text{test}} \rightarrow A_{\text{analysis}}.
\label{eq:pipeline}
\end{equation}

The Requirement Agent $A_{\text{req}}$ generates system and software requirements and seeds \texttt{REF\_BY} links with derivation confidence $c_{\text{seed}}$. The Architect Agent $A_{\text{arch}}$ generates software components and architectural elements that implement system functionality. The Traceability Agent $A_{\text{trace}}$ validates seeded links and creates new traceability relationships across all artifact types using the two-stage pipeline described in Section~\ref{sec:twostage}. The Test Agent $A_{\text{test}}$ generates test specifications and establishes \texttt{VER\_BY} relationships. Finally, the Analysis Agent $A_{\text{analysis}}$ executes graph queries and analytics to generate project-level reports on traceability coverage, gaps, and conflicts.

All agents interact with $\mathcal{G}$ through the centralized Graph Service, ensuring a consistent global context across the pipeline.

\subsection{Two-Stage Traceability Link Prediction}
\label{sec:twostage}

Traceability links are generated through a two-stage process combining semantic retrieval with LLM-based multi-criteria reasoning.

\subsubsection{Stage 1: Semantic Retrieval}

Each artifact is embedded using a pre-trained sentence encoder $\mathcal{M}_{\text{enc}}$: $\mathbf{e}_v = \mathcal{M}_{\text{enc}}(\text{desc}(v))$. For a given requirement $r$, candidate artifacts are retrieved using cosine similarity:
\begin{equation}
\text{sim}(r, c) = \frac{\mathbf{e}_r \cdot \mathbf{e}_c}{\|\mathbf{e}_r\| \, \|\mathbf{e}_c\|},
\label{eq:cosine}
\end{equation}
\begin{equation}
\mathcal{C}_r = \underset{\mathcal{S} \subseteq \mathcal{C}, |\mathcal{S}| = k}{\arg\max} \sum_{c \in \mathcal{S}} \text{sim}(r, c).
\label{eq:topk}
\end{equation}
This narrows the search space from $|\mathcal{C}|$ to $k$ candidates where $k \ll |\mathcal{C}|$.

\subsubsection{Stage 2: LLM-Based Multi-Criteria Analysis}

For each candidate pair $(r, c)$ with $c \in \mathcal{C}_r$, a structured prompt $P(r, c)$ directs the LLM to evaluate four reasoning criteria jointly: \textit{requirement coverage}, assessing whether the component fulfills the functional intent of the requirement; \textit{implementation specificity}, assessing whether the component description indicates concrete related functionality; \textit{terminology alignment}, assessing whether the artifacts share meaningful domain terminology; and \textit{absence of contradiction}, assessing whether the component suggests unrelated or conflicting responsibilities.

The LLM $\mathcal{M}_{\text{LLM}}$ jointly evaluates these criteria and produces:
\begin{equation}
(y_{rc}, c_{rc}, j_{rc}) = \mathcal{M}_{\text{LLM}}(P(r, c)),
\end{equation}
where $y_{rc} \in \{0, 1\}$ is the link decision, $c_{rc} \in [0, 1]$ is the confidence score, and $j_{rc}$ is a natural language justification. Confidence values are interpreted using structured bands: 0.90--1.00 indicates clear and direct implementation, 0.70--0.89 strong implementation, 0.50--0.69 partial implementation, 0.30--0.49 indirect relevance, and 0.00--0.29 no meaningful relationship.

When a valid link is identified, the corresponding edge is created in $\mathcal{G}$ carrying the confidence score, justification, and creating agent identifier as properties.

\subsection{Traceability Seeding and Validation}

\subsubsection{Derivation-Time Seeding}

When $A_{\text{req}}$ derives a software requirement $s$ from a system requirement $r$, it creates a \texttt{REF\_BY} edge with a self-assessed derivation confidence:
\begin{equation}
e_{\text{seed}} = (s, r), \quad c(e_{\text{seed}}) = c_{\text{seed}}, \quad s(e_{\text{seed}}) = \texttt{PENDING}.
\label{eq:seed}
\end{equation}

This seeding provides the Traceability Agent with a warm start: rather than treating all artifact pairs as equally unknown, $A_{\text{trace}}$ enters its validation pass with structurally motivated links whose confidence values serve as derivation-time priors.

\subsubsection{Global Validation Pass}

After all derivation turns are complete, $A_{\text{trace}}$ executes a global two-stage traceability pass over the full Cartesian product of system and software requirements, not restricted to seeded pairs. For each evaluated pair, the pipeline produces a validation confidence $c_{\text{val}}$. When a seeded link already exists, $c_{\text{val}}$ is recorded alongside $c_{\text{seed}}$ on the existing edge:
\begin{equation}
e_{\text{seed}}.\{c_{\text{val}}, \alpha_{\text{val}}, \tau_{\text{val}}\} \leftarrow (c_{\text{val}}, A_{\text{trace}}, \tau_{\text{now}}).
\label{eq:val}
\end{equation}

When a valid relationship is identified for a pair without an existing link, a new edge is created with $\alpha = A_{\text{trace}}$. This ensures the graph captures both provenance-based links recording structural derivation origins and inference-based links representing semantically valid relationships that may cross derivation boundaries.

\subsection{Conflict Detection}

The co-existence of derivation-time and validation-time assessments enables structured conflict detection. Two categories of conflict are identified.

\subsubsection{Confidence Divergence Conflict}

A confidence divergence conflict arises when the gap between seeded and validated confidence for the same edge exceeds a threshold $\delta$:
\begin{equation}
\Delta c = c_{\text{seed}} - c_{\text{val}}, \quad \text{conflict triggered if } |\Delta c| > \delta.
\label{eq:divergence}
\end{equation}

This indicates that the Requirement Agent's self-assessment disagrees with the Traceability Agent's independent evaluation when the software requirement is considered in the context of all system requirements simultaneously. The divergence may reflect semantic drift introduced during derivation. The severity of the conflict is set to $|\Delta c|$.

\subsubsection{Cross-Cluster Ambiguity Conflict}

System requirements are grouped into semantic clusters based on embedding similarity. Let $C(r)$ denote the cluster assignment of requirement $r$ and $d(C_i, C_j)$ the inter-cluster semantic distance. A cross-cluster ambiguity conflict is triggered when a software requirement $s$ holds valid \texttt{REF\_BY} links to system requirements in semantically distant clusters:
\begin{equation}
c_{\text{val}}(s, r_i) \geq \tau \;\land\; c_{\text{val}}(s, r_j) \geq \tau \;\land\; d(C(r_i), C(r_j)) > \epsilon,
\label{eq:crosscluster}
\end{equation}
where $\tau$ is the acceptance confidence threshold and $\epsilon$ is the inter-cluster distance threshold. This condition signals that the software requirement serves two unrelated system concerns, typically indicating a need for decomposition.

\subsubsection{Conflict Materialization}

\begin{figure}[h]
  \centering
  \includegraphics[width=0.9\linewidth]{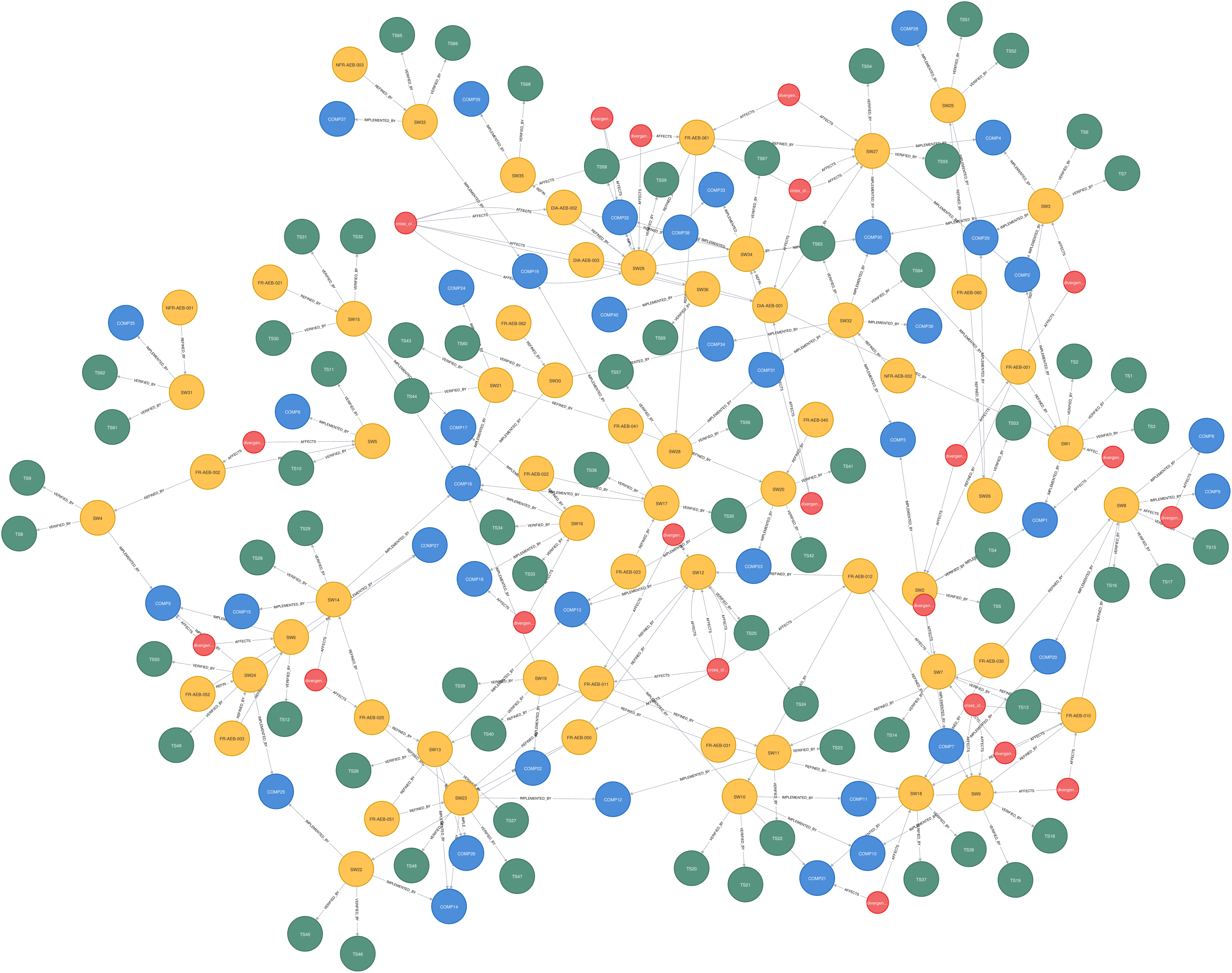}
  \caption{Complete Knowledge Graph Schema. Artifact nodes (SysReq, SwReq, Comp, Test) linked by Refinement,
  Implementation, and Verification relationships. Conflict nodes (DIVERGENCE, CROSS\_CLUSTER) materialize inconsistencies
  detected during validation, connected via AFFECTS edges.}
  \label{fig:kg_schema}
  \end{figure}

Both conflict types are materialized as \texttt{Conflict} nodes in $\mathcal{G}$, as illustrated in Fig.~\ref{fig:kg_schema}, with properties including type $\in \{\texttt{DIVERGENCE}, \texttt{CROSS\_CLUSTER}\}$, severity $\in [0, 1]$, and resolution status $\in \{\texttt{OPEN}, \texttt{RESOLVED}, \texttt{ACCEPTED\_RISK}\}$. Each conflict node is connected to affected artifacts via \texttt{AFFECTS} edges. By representing conflicts as graph entities rather than external logs, the system allows engineers and the Analysis Agent to perform conflict resolution within the same unified graph context.

\subsection{Graph-Based Engineering Analysis}

The graph structure enables the Analysis Agent $A_{\text{analysis}}$ to perform several engineering analyses through Cypher queries and graph algorithms:

\noindent\textbf{Impact Analysis.} Graph traversal determines which components and tests are affected by changes to a requirement, computed as the set of nodes reachable within $h$ hops:
\begin{equation}
\text{Impact}(v, h) = \{u \in \mathcal{V} \mid \text{dist}(v, u) \leq h\}.
\label{eq:impact}
\end{equation}

\noindent\textbf{Traceability Gap Detection.} Missing links are identified as requirements without outgoing edges of expected types:
\begin{equation}
\text{Gaps}_{\texttt{IMP}} = \{r \in \mathcal{R} \mid \nexists \, c \in \mathcal{C}: (r, c) \in \mathcal{E}_{\texttt{IMP\_BY}}\}.
\label{eq:gaps}
\end{equation}

\noindent\textbf{Test Coverage.} Coverage is computed as the ratio of requirements with at least one \texttt{VER\_BY} link:
\begin{equation}
\text{Coverage} = \frac{|\{r \in \mathcal{R} \mid \exists \, t \in \mathcal{T}: (r, t) \in \mathcal{E}_{\texttt{VER\_BY}}\}|}{|\mathcal{R}|}.
\label{eq:coverage}
\end{equation}

\noindent\textbf{Centrality Analysis.} Degree centrality identifies critical components that implement the largest number of requirements:
\begin{equation}
\text{centrality}(c) = |\{r \in \mathcal{R} \mid (r, c) \in \mathcal{E}_{\texttt{IMP\_BY}}\}|.
\label{eq:centrality}
\end{equation}

\noindent\textbf{Requirement Clustering.} Semantically related requirements are grouped using embedding similarity to reveal functional domains, subsystem boundaries, and hidden dependencies. The resulting clusters support the cross-cluster ambiguity detection described in Eq.~\ref{eq:crosscluster}.

The Analysis Agent executes these queries to generate project-level reports summarizing implementation coverage, test verification status, traceability completeness, open conflicts, and dependency insights. These reports provide engineers with an up-to-date overview and surface gaps or inconsistencies for resolution.

\section{Experiments}

\subsection{Dataset Description}

To evaluate the proposed multi-agent framework, we compiled a dataset of 535 software and system artifacts from six automotive subsystems: Autonomous Emergency Braking (AEB), ADAS Lane Keeping (LKA), EV Battery Management (BMS), Vehicle Infotainment (IVI), V2X Communication, and Driver Monitoring (DMS). The artifacts span four categories: System Requirements (83), Software Requirements (128), Architectural Components (124), and Test Specifications (200).

The dataset is structured following the automotive V-cycle development model, capturing three primary semantic relationships: (1) \emph{Refinement} linking System Requirements to Software Requirements, (2) \emph{Implementation} linking Software Requirements to Components, and (3) \emph{Verification} linking Requirements and Components to Test Specifications. Human-validated traceability matrices were established for all six subsystems prior to experiments, providing ground truth against which AI-predicted links are evaluated.
 \subsection{Experiment 1: Two-Stage Pipeline Validation}

  \textbf{Objective:} Validate that the two-stage pipeline (semantic retrieval + LLM reasoning) outperforms
  industry-standard approaches for traceability link prediction across all V-cycle relationships (\texttt{REF\_BY},
  \texttt{IMP\_BY}, \texttt{VER\_BY}).

  \textbf{Baselines:} (1) TF-IDF + Cosine (lexical matching), (2) Embedding-only (dense embeddings, Stage 1 alone), (3)
  LLM-only (unfiltered LLM evaluation), (4) Full Pipeline (proposed). Metrics: Precision, Recall, F1-Score, MAP, evaluated
   against human-validated ground truth.

  \textbf{Results:} Table~\ref{tab:exp1_results} shows that the Full Pipeline achieved \textbf{F1 = 0.769}, substantially
  outperforming TF-IDF (F1 = 0.349, lexical matching only) and LLM-only (F1 = 0.669, context-window saturation). The
  two-stage approach recovers 13\% F1 improvement by pre-filtering the Cartesian product (200+ pairs) down to $k \approx
  10$ high-confidence candidates before LLM reasoning.

  \textbf{Why This Matters:} A 13\% F1 improvement translates directly to compliance impact: at this recall rate (0.687),
  fewer than 1 in 3 valid requirements are missed, meeting ASPICE/ISO 26262 traceability completeness requirements.
  TF-IDF's 0.228 recall would leave 77\% of requirements orphaned—unacceptable in safety-critical domains. LLM-only's
  0.699 recall, though reasonable, still misses 30\% of valid links due to processing 200+ pairs simultaneously without
  semantic pre-filtering.

  \begin{table}[h]
        \centering
        \footnotesize
        \caption{Traceability Link Prediction Accuracy (Global Average)}
        \label{tab:exp1_results}
        \vspace{-0.2cm}
        \begin{tabular}{|l|c|c|c|c|}
                \hline
                \textbf{Baseline} & \textbf{Precision} & \textbf{Recall} & \textbf{F1-Score} & \textbf{MAP} \\
                \hline
                TF-IDF + Cosine & 1.000 & 0.228 & 0.349 & 0.894 \\
                Embedding-only & 0.829 & 0.637 & 0.707 & 0.913 \\
                LLM-only & 0.705 & 0.699 & 0.669 & 0.796 \\
                \textbf{Full Pipeline} & \textbf{0.900} & \textbf{0.687} & \textbf{0.769} & \textbf{0.905} \\
                \hline
        \end{tabular}
  \end{table}

  \subsubsection*{Performance by Traceability Relationship}

  Table~\ref{tab:exp1_relationships} reveals relationship-specific performance. \texttt{REF\_BY} (System $\to$
  Software Req) achieves \textbf{perfect precision (1.000)} but moderate recall due to the large abstraction gap---the
  pipeline is appropriately conservative at this boundary, ensuring no false positive derivations.
  \texttt{IMP\_BY} (Software Req $\to$ Component) shows \textbf{best balance (F1 = 0.833, MAP = 0.939)} with
  shared technical vocabulary---this close domain match enables highest confidence in AI-predicted links.
  \texttt{VER\_BY} (Software Req $\to$ Test) scores lowest (F1 = 0.706) due to behavioral vs. functional vocabulary
  mismatch, revealing where human expert review is still essential.

  \begin{table}[h]
        \centering
        \footnotesize
        \caption{Full Pipeline Accuracy by Relationship Type}
        \label{tab:exp1_relationships}
        \vspace{-0.2cm}
        \begin{tabular}{|l|l|c|c|c|}
                \hline
                \textbf{Relationship} & \textbf{Path} & \textbf{Precision} & \textbf{F1} & \textbf{MAP} \\
                \hline
                \texttt{REF\_BY} & SysReq$\to$SwReq & 1.000 & 0.842 & 0.875 \\
                \texttt{IMP\_BY} & SwReq$\to$Comp & 0.833 & 0.833 & 0.939 \\
                \texttt{VER\_BY} & SwReq$\to$TestSpec & 0.857 & 0.706 & 0.773 \\
                \hline
        \end{tabular}
  \end{table}

\subsection{Experiment 2: Conflict Detection Parameter Ablation}

\textbf{Objective:} Empirically optimize conflict detection \{\texttt{DIVERGENCE}, \texttt{CROSS\_CLUSTER}\} by performing a targeted ablation
study on cross-cluster ambiguity conflicts to tune the hyperparameters ($\tau$, $\varepsilon$),
while establishing and justifying a fixed setting for the confidence divergence threshold
$\delta$, ensuring the system identifies requirements that violate architectural constraints
while maintaining zero false positives.
\textbf{Setup:} The conflict detection mechanism employs three hyperparameters:
\begin{enumerate}
    \item\texttt{DIVERGENCE} confidence threshold $\delta=0.5$, a qualitative domain-expert parameter
representing tolerance for semantic drift between requirement derivation and
validation---held constant and not ablated;
\item\texttt{CROSS\_CLUSTER} confidence threshold $\tau$, the minimum confidence for conflict flagging;
\item\texttt{CROSS\_CLUSTER} inter-cluster distance threshold $\varepsilon$, the semantic distance upper bound for considering clusters related.
\end{enumerate}

We perform a systematic ablation sweep over
$\tau \in \{0.3, 0.5, 0.6, 0.8\}$ and $\varepsilon \in \{0.3, 0.4, 0.5, 0.6, 0.7\}$
(120 configurations) across four conflict-bearing datasets (LKA, DMS, AEB, EV) with 10
logically valid conflicts validated by human review.
Metrics: True Positives (TP) and False Negatives (FN).

\textbf{Results:} Table~\ref{tab:exp2_ablation} presents the ablation grid across four
datasets.
The configuration $(\tau=0.5,\,\varepsilon=0.4)$ uniquely achieves TP=10, FN=0, detecting
all ground-truth conflicts.
Critically, zero false positives were generated across all 120 tested configurations,
demonstrating conservative detection by design.

The ablation reveals critical dataset-dependent parameter dominance.
DMS conflicts consistently suppress at $\varepsilon \geq 0.5$ (cluster distance binding),
insensitive to $\tau$ within the safe range---this explains the vertical failure pattern in
the DMS column (Table~\ref{tab:exp2_ablation}).
In contrast, EV conflicts exhibit $\tau$-dominance, suppressing at $\tau \geq 0.7$ with no
distance-based suppression in the tested range (visible as the diagonal failure pattern at
$\tau=0.8$).
LKA and AEB show mixed behaviour: some conflicts respond to either parameter alone, while
others require both in tandem.
This heterogeneity is fundamental: no single-parameter adjustment can achieve global
optimality.
Increasing $\varepsilon$ alone misses DMS; increasing $\tau$ alone misses EV; only the
joint constraint $(\tau=0.5,\,\varepsilon=0.4)$ satisfies all per-conflict suppression
thresholds simultaneously, as verified in Table~\ref{tab:exp2_ablation}.

\textbf{Parameter Selection Rationale:}
\begin{itemize}
    \item $\tau=0.5$ represents the principled minimum confidence below which a
    \texttt{REF\_BY} relation cannot be considered reliably indicative of a true
    derivation link---a sub-$0.5$ confidence implies the relation is more uncertain than
    certain, providing no logical basis for flagging a conflict. All 10 ground-truth
    conflicts carry minimum relation confidences between $0.67$ and $0.81$, well above
    this boundary.
    \item $\varepsilon=0.4$ is anchored in the geometric interpretation of inter-cluster
    cosine distance: clusters separated by less than $0.4$ share substantial functional
    overlap and cannot be considered architecturally distinct, making conflict detection
    between them logically incoherent. It is the tightest safe value that captures all
    valid cross-cluster conflicts without collapsing semantically distinct clusters.
\end{itemize}

Taken together, $(\tau=0.5,\,\varepsilon=0.4)$ is the most conservative configuration
within the perfect-detection zone: any relaxation admits logically unjustified signals;
any tightening risks missing domain-specific conflicts, as confirmed by
Table~\ref{tab:exp2_ablation}.

\begin{table}[ht]
\centering
\caption{Conflict Detection Ablation Results ($\tau \times \varepsilon$ Grid;
TP/FN = True Positives / False Negatives)}
\label{tab:exp2_ablation}
\vspace{-0.2cm}
\footnotesize
\begin{tabular}{|c|c|c|c|c|c|c|}
\hline
\textbf{$\tau$} & \textbf{$\varepsilon$} &
\textbf{LKA} & \textbf{DMS} & \textbf{AEB} & \textbf{EV} & \textbf{Comb.} \\
\hline
0.3 & 0.3 & 2/0 & 3/0 & 2/0 & 3/0 & 10/0 \\
    & 0.4 & 2/0 & 3/0 & 2/0 & 3/0 & 10/0 \\
    & 0.5 & 2/0 & 0/3 & 2/0 & 3/0 & 7/3  \\
    & 0.6 & 2/0 & 0/3 & 0/2 & 3/0 & 5/5  \\
    & 0.7 & 0/2 & 0/3 & 0/2 & 2/1 & 2/8  \\
\hline
\textbf{0.5} & 0.3 & 2/0 & 3/0 & 2/0 & 3/0 & 10/0 \\
 & \textbf{0.4}$\star$ & \textbf{2/0} & \textbf{3/0} & \textbf{2/0} & \textbf{3/0} & \textbf{10/0} \\
 & 0.5 & 2/0 & 0/3 & 2/0 & 3/0 & 7/3  \\
 & 0.6 & 2/0 & 0/3 & 0/2 & 3/0 & 5/5  \\
 & 0.7 & 0/2 & 0/3 & 0/2 & 2/1 & 2/8  \\
\hline
0.8 & 0.3 & 0/2 & 1/2 & 0/2 & 0/3 & 1/9  \\
    & 0.4 & 0/2 & 1/2 & 0/2 & 0/3 & 1/9  \\
    & 0.5 & 0/2 & 0/3 & 0/2 & 0/3 & 0/10 \\
    & 0.6 & 0/2 & 0/3 & 0/2 & 0/3 & 0/10 \\
    & 0.7 & 0/2 & 0/3 & 0/2 & 0/3 & 0/10 \\
\hline
\end{tabular}
\vspace{0.1cm}
\end{table}

\subsection{Experiment 3: Consistency Control Ablation}

\textbf{Objective:} Evaluate the impact of downstream consistency controls (confidence filtering, divergence conflict detection, and cross-cluster ambiguity detection) on the quality and usability of the traceability graph. The experiment analyzes the trade-off between trace-link growth and structural consistency in the knowledge graph.

\textbf{Setup:} A seeded knowledge graph is constructed for each dataset, and the downstream pipeline (Traceability $\rightarrow$ Cross-Cluster Detection $\rightarrow$ PULSE) is executed under four ablation conditions: (1) No Consistency Controls, (2) No Confidence Filtering, (3) No Conflict Detection, and (4) Full Consistency Stack. Metrics are extracted from the final KG state. Results are averaged over five datasets (LKA, AEB, BMS, V2X, IVI).

Metrics are derived from the final KG state and cover both traceability growth (final and new links, inflation ratio) and consistency aspects (conflict rate and conflict-aware path completeness), reflecting the trade-off between expansion and structural soundness.

\textbf{Results:} Tables~\ref{tab:exp3_growth} and \ref{tab:exp3_conflicts} summarize the results. The findings reveal a clear trade-off between graph expansion and consistency. Conditions without consistency controls produce larger graphs but either hide or introduce ambiguity, while the Full Consistency Stack produces a more controlled graph and explicitly exposes structural inconsistencies, leading to more realistic traceability evaluation.

\begin{table}[H]
\centering
\footnotesize
\caption{Traceability Growth Metrics (Averaged over 5 Datasets)}
\label{tab:exp3_growth}
\vspace{-0.2cm}
\begin{tabular*}{0.5\textwidth}{@{\extracolsep{\fill}}|l|c|c|c|}
\hline
\textbf{Condition} & \textbf{Final Links} & \textbf{New Links} & \textbf{Inflation Ratio} \\
\hline
No Consistency Controls & 169.4 & 126.4 & 3.94 \\
No Confidence Filtering & 147.6 & 104.6 & 3.43 \\
No Conflict Detection & 65.6 & 22.6 & 1.48 \\
\textbf{Full Consistency Stack} & \textbf{64.4} & \textbf{21.4} & \textbf{1.46} \\
\hline
\end{tabular*}
\end{table}

\begin{table}[H]
\centering
\footnotesize
\caption{Conflict and Usability Metrics (Averaged over 5 Datasets)}
\label{tab:exp3_conflicts}
\vspace{-0.2cm}
\setlength{\tabcolsep}{2pt}
\begin{tabular}{|l|c|c|c|}
\hline
\textbf{Condition} & \textbf{Conflict} & \textbf{Mean} & \textbf{Path Comp.} \\
 & \textbf{Rate (\%)} & \textbf{Severity} & \textbf{(\%)} \\
\hline
No Consistency Controls & 0.00 & 0.00 & 100.00 \\
No Confidence Filtering & 3.97 & 0.6569 & 36.12 \\
No Conflict Detection & 0.00 & 0.00 & 100.00 \\
\textbf{Full Consistency Stack} & \textbf{15.42} & \textbf{0.5207} & \textbf{33.64} \\
\hline
\end{tabular}
\end{table}

\subsection{Experiment 4: Link Creation Threshold Ablation}

\textbf{Objective:} Empirically optimize the link creation threshold ($\theta_{create}$) to maximize the accuracy of the automated Knowledge Graph while minimizing the manual verification load.

\textbf{Setup:} We performed a systematic ablation sweep over $\theta_{create} \in [0.0, 0.9]$ across five automotive datasets. For each threshold, we measured the average Precision, Recall, and F1-Score of the predicted traceability links against human-validated ground truth. Additionally, we tracked the \emph{Queue Size}, representing the average number of links per dataset falling between $\theta_{create}$ and the auto-confirmation threshold, thus requiring manual expert review.

\textbf{Results:} Table~\ref{tab:exp4_thresholds} illustrates the sensitivity of the traceability graph to the creation threshold. At low thresholds ($\theta \leq 0.4$), the system achieves high recall ($\approx 0.90$) but suffers from low precision ($\approx 0.28$--$0.33$) and a prohibitive manual verification load ($> 90$ links per dataset). As the threshold increases, we observe a sharp improvement in Precision. The configuration $\theta_{create} = 0.7$ emerges as the optimal balance, yielding a peak average F1-Score of $0.7958$ with a manageable manual queue of $23.0$ links. Beyond $\theta = 0.8$, the gain in Precision is offset by a precipitous drop in Recall, as the agent becomes overly conservative and misses valid semantic connections.

\begin{table}[h]
\centering
\footnotesize
\caption{Impact of Creation Threshold ($\theta_{create}$) on Traceability Accuracy and Manual Effort (Averaged across 5 Datasets)}
\label{tab:exp4_thresholds}
\vspace{-0.2cm}
\begin{tabular}{|c|c|c|c|c|}
\hline
\textbf{$\theta_{create}$} & \textbf{Precision} & \textbf{Recall} & \textbf{F1-Score} & \textbf{Queue Size} \\
\hline
0.0 & 0.2846 & 0.9019 & 0.4327 & 112.4 \\
0.1 & 0.2846 & 0.9019 & 0.4327 & 112.4 \\
0.2 & 0.2951 & 0.9019 & 0.4443 & 108.6 \\
0.3 & 0.3268 & 0.9019 & 0.4787 & 96.6 \\
0.4 & 0.3357 & 0.8968 & 0.4872 & 93.0 \\
0.5 & 0.4610 & 0.8915 & 0.6037 & 63.0 \\
0.6 & 0.6121 & 0.8811 & 0.7150 & 43.2 \\
\textbf{0.7} & \textbf{0.7815} & \textbf{0.8183} & \textbf{0.7958} & \textbf{23.0} \\
0.8 & 0.9249 & 0.6506 & 0.7578 & 7.0 \\
0.9 & 0.9826 & 0.3397 & 0.5004 & 0.0 \\
\hline
\end{tabular}
\end{table}
\section{Conclusion}

This paper addressed how multiple AI agents can reliably collaborate over shared software artifacts in safety-critical domains. We demonstrated that confidence-calibrated traceability links serve as both prediction mechanisms and coordination signals, enabling downstream agents to assess upstream reliability and detect inconsistencies through divergence detection. Our four-experiment evaluation on 535 automotive artifacts established that the two-stage pipeline achieves F1=0.769, outperforming lexical (F1=0.349) and unfiltered LLM approaches (F1=0.669); that hyperparameter optimization is dataset-dependent, with $(\tau=0.5, \varepsilon=0.4)$ uniquely achieving perfect detection (10/10) with zero false positives; that the full consistency stack controls graph inflation (inflation ratio 1.46 vs. 3.94 without controls) while exposing structural inconsistencies; and that an optimal link creation threshold $\theta_{create}=0.7$ balances precision (0.782) and recall (0.818) with a manageable manual effort of 23 links per dataset. These results collectively confirm that calibrated confidence is essential for effective pipeline coordination, and that treating the shared knowledge graph as an active coordination surface enables the framework to address ASPICE and ISO~26262 compliance by providing an auditable record of inconsistency detection and resolution.

At the same time, the evaluation is limited to automotive systems (535 artifacts across 6 subsystems) within safety-critical regulatory contexts, and generalization to larger codebases as well as the bootstrap problem of confidence calibration in domains lacking ground truth remain open. The domain-expert parameter $\delta$ is manually set rather than empirically tuned, reflecting a deliberate preservation of human authority over semantic tolerance. Building on these limitations, promising directions for future work include automated conflict resolution, domain generalization to adjacent regulated areas such as avionics and medical devices, and confidence calibration in the absence of ground truth. More broadly, the pattern of using calibrated confidence as a coordination signal generalizes to any multi-stage AI pipeline in which upstream decisions constrain downstream behavior, and the core contribution---that \emph{confidence enables trust-aware coordination in multi-agent systems}---is particularly timely as AI agents assume increasingly critical roles in regulated software engineering.

\bibliographystyle{IEEEtran}
\bibliography{ref}
\end{document}